\magnification=1200
\baselineskip=14pt

\def\P{\Phi}
\def\L{\Lambda}
\def\a{\alpha}
\def\b{\beta}

\def\d{\delta}
\def\l{\lambda}
\def\N{{\cal N}}
\def\G{{\cal G}}

\rightline{UCLA/99/TEP/3}
\rightline{Columbia/99/Math}

\bigskip
\centerline{\bf LAX PAIRS AND
SPECTRAL CURVES FOR CALOGERO-MOSER}
\centerline{{\bf AND SPIN CALOGERO-MOSER SYSTEMS}
\footnote*{Research supported in part by the
National Science Foundation under grants PHY-95-31023
and DMS-98-00783.}}

\bigskip

\centerline{\bf Eric D'Hoker${}^1$ and D.H. Phong${}^2$}

\bigskip

\centerline{${}^1$ Department of Physics}
\centerline{University of California, Los Angeles, CA 90024}
\centerline{Institute for Theoretical Physics}
\centerline{University of California, Santa Barbara, CA 93106}
\bigskip

\centerline{${}^2$ Department of Mathematics}
\centerline{Columbia University, New York, NY 10027}

\bigskip

\centerline{\bf Abstract}

\bigskip

We summarize recent results on the construction of Lax pairs
with spectral parameter for the twisted and untwisted elliptic
Calogero-Moser systems associated with arbitrary simple
Lie algebras, their scaling limits to Toda systems,
and their role in Seiberg-Witten theory.
We extend part of this work by presenting a new
parametrization for the spectral curves for elliptic
spin Calogero-Moser systems associated with $SL(N)$.

\vskip 1in

\centerline{{\it Contribution to the issue of ``Regular and Chaotic Dynamics"}}
\centerline{{\it dedicated to Professor J. Moser}}
\centerline{{\it on the occasion of his 70-th birthday}}

\vfill\break

\centerline{\bf I. INTRODUCTION}

\bigskip

Calogero-Moser systems are Hamiltonian systems with an
amazingly rich structure. Recently, another remarkable
property of these systems has been brought to light,
namely their intimate connection with exact solutions
of supersymmetric gauge theories.

\medskip
The $\N=2$ supersymmetric $SU(N)$ gauge theory
with a hypermultiplet in the adjoint
representation was the first gauge theory to
be linked with elliptic Calogero-Moser systems.
In their 1995 work, based on several
consistency checks, Donagi and Witten [1] had proposed
that the Seiberg-Witten spectral curves for the
low-energy exact solution of this theory were given
by the spectral curves of a $SU(N)$ Hitchin system.
Krichever in unpublished work, Gorsky and Nekrasov and
Martinec [2] have
recognized the $SU(N)$ Hitchin spectral
curves as identical to the spectral curves for 
elliptic $SU(N)$ Calogero-Moser
systems. That the $SU(N)$ elliptic Calogero-Moser curves
do provide the Seiberg-Witten
solution of the $SU(N)$ gauge theory with matter in the
adjoint representation was established by the authors in
[3]. In particular, it was shown in [3] that
the resulting prepotential has the correct logarithmic
singularities predicted by field theoretic
perturbative calculations, and that it satisfies
a renormalization group equation which 
determines explicitly instanton contributions to any order.

\medskip
The major problem in Seiberg-Witten theory
is to determine the spectral curves, and hence the
integrable models, corresponding to an arbitrary
asymptotically free or conformally
invariant $\N=2$ supersymmetric gauge theory with gauge algebra $\G$,
and matter hypermultiplets in
a representation $R$ of $\G$.
For reviews, see e.g. [4][5,6,7].
It has been known now for a long time,
thanks to the work of Olshanetsky and Perelomov [8][9],
that Calogero-Moser systems can be defined for any simple
Lie algebra\footnote*{Other models associated to
Lie algebras include the Toda systems, of which more will
be said below, and the Ruijsenaars-Schneider systems,
whose role in gauge theories is still obscure.}.
Olshanetsky and Perelomov also showed that
the Calogero-Moser systems for classical Lie
algebras were integrable,
although the existence of a spectral curve
(or Lax pair with spectral parameter) as well as the case
of exceptional Lie algebras remained open.
Thus several immediate questions were:

\medskip

$\bullet$ Does the elliptic Calogero-Moser system for general
Lie algebra $\G$ admit a Lax pair with spectral parameter? 

$\bullet$ Does it correspond to the $\N=2$ supersymmetric gauge
theory with gauge algebra $\G$ and a hypermultiplet in
the adjoint representation?

$\bullet$ Can this correspondence be verified in the limiting
cases when the mass $m$ of the hypermultiplet tends to
$0$ and the gauge theory acquires an $\N=4$ supersymmetry and
becomes exact, and in the limit $m\rightarrow\infty$, when the hypermultiplet
decouples and the theory reduces to pure $\N=2$ Yang-Mills?

\medskip
 
The answers to these questions turn out to
be the following [10,11,12].

\bigskip

$\bullet$ The elliptic Calogero-Moser systems defined
by an arbitrary simple Lie algebra $\G$ do admit Lax
pairs with spectral parameters.
(In the case of $E_8$, we need to assume the existence
of a cocycle)[10].

$\bullet$ The correspondence between elliptic $\G$ Calogero-Moser
systems and $\N=2$ supersymmetric $\G$ gauge theories
with matter in the adjoint representation
is only correct when
the Lie algebra $\G$ is simply-laced.
When $\G$ not simply-laced, we require 
new integrable models, namely the
twisted elliptic Calogero-Moser systems
introduced in [10,12]. 

$\bullet$ The new twisted elliptic Calogero-Moser systems
also admit a Lax pair with spectral parameter,
except possibly in the case $\G=G_2$ [10].

$\bullet$ In the scaling limit $m=Mq^{-{1\over 2}\d}\rightarrow\infty$,
$M$ fixed, the twisted (respectively untwisted) elliptic $\G$
Calogero-Moser
systems tend to the Toda system for
$(\G^{(1)})^{\vee}$ (respectively $\G^{(1)}$)
for $\d={1\over h_{\G}^{\vee}}$ (respectively $\d={1\over h_{\G}}$).
Here $h_{\G}$ and $h_{\G}^{\vee}$ are the Coxeter and the dual Coxeter numbers
of $\G$ [11]. 

\bigskip

The main purpose of this paper is to review some
of these developments. 
Although the case of the adjoint representation
for arbitrary gauge algebras has now been solved,
the correspondence between gauge theories and
integrable models is still far from complete.
In particular, one can wonder about the
eventual role, if any, of other generalizations
of elliptic Calogero-Moser systems
such as the Ruijsenaars-Schneider systems [13,14]
or the spin Calogero-Moser systems [15].
Such questions require a better understanding of
the spectral curves of these systems, and
particularly of their parametrizations.
Thus we have taken this opportunity to
describe also a new parametrization
for the spectral curves of spin Calogero-Moser
systems. This new parametrization is suggestive of
the order parameters for the $SU(N)$ gauge theory,
and may be valuable in future developments.
See also [30] for recent developments.

\bigskip
 
\centerline{\bf II. TWISTED AND UNTWISTED CALOGERO-MOSER SYSTEMS}

\bigskip

\noindent
{\bf The $SU(N)$ Elliptic Calogero-Moser System}

The basic system in this paper is the elliptic Calogero-Moser
system defined by the Hamiltonian
$$
H(x,p)={1\over 2}\sum_{i=1}^Np_i^2-{1\over 2}m^2
\sum_{i\not=j}\wp(x_i-x_j)
\eqno(2.1)
$$
Here $m$ is a mass parameter,
and $\wp(z)$ is the Weierstrass $\wp$-function, defined on a torus
${\bf C}/(2\omega_1{\bf Z}+2\omega_2{\bf Z})$. As usual, we denote by
$\tau=\omega_2/\omega_1$ the moduli of the torus, and set $q=e^{2\pi i\tau}$.
The well-known trigonometric and
rational limits with respective potentials 
$$
-{1\over 2}m^2\sum_{i\not=j}
{1\over 4\,{\rm sh}^2\,{x_i-x_j\over 2}}
\qquad
{\rm and}
\qquad 
-{1\over 2}m^2\sum_{i\not=j}{1\over (x_i-x_j)^2}
$$ 
arise
in the limits $\omega_1=-i\pi,\omega_2\rightarrow\infty$
and $\omega_1,\omega_2\rightarrow\infty$.
All these systems have been shown to be completely integrable
in the sense of Liouville, i.e. they all admit a complete set of integrals of motion
which are in involution [16,17].

\bigskip

Our considerations require however a notion of integrability which
is in some sense more stringent, namely a Lax pair $L(z)$, $M(z)$
with spectral parameter $z$. Such a Lax pair was obtained by Krichever [18]
in 1980. He showed that the Hamiltonian system (2.1) is
equivalent to the Lax equation $\dot L(z)=[L(z),M(z)]$,
with $L(z)$ and $M(z)$ given by the following $N\times N$
matrices
$$
\eqalignno{
L_{ij}(z)=&p_i\d_{ij}-m(1-\d_{ij})\P(x_i-x_j,z)\cr
M_{ij}(z)=&m\d_{ij}\sum_{k\not= i}\wp(x_i-x_k)-m(1-\d_{ij})\P'(x_i-x_j,z).
&(2.2)\cr}
$$ 
The function $\P(x,z)$ is defined by
$$
\P(x,z)={\sigma(z-x)\over
\sigma(z)\sigma(x)}e^{x\zeta(z)},
\eqno(2.3)
$$
where $\sigma(z)$, $\zeta(z)$ are the usual Weierstrass $\sigma$
and $\zeta$ functions on the torus ${\bf C}/(2\omega_1{\bf Z}+2\omega_2{\bf Z})$.
The function $\P(x,z)$ satisfies the key functional equation
$$
\Phi(x,z)\P'(y,z)
-\P(y,z)\P'(x,z)
=(\wp(x)-\wp(y))\P(x+y,z).
\eqno(2.4)
$$ 
It is well-known that
functional equations of this form 
are required for the Hamilton
equations of motion to be equivalent to
the Lax equation
$\dot L(z)=[L(z),M(z)]$
with a Lax pair of the form (2.2).
Often, solutions had been obtained under
additional parity assumptions
in $x$ (and $y$), which prevent
the existence of a spectral parameter. 
The solution $\P(x,z)$
with spectral parameter $z$
is obtained by dropping such parity assumptions
for general $z$. It is a relatively recent result
of Braden and Buchstaber [19] that, conversely,
the functional equation (2.4) essentially
determines $\P(x,z)$.

\bigskip

\noindent
{\bf Calogero-Moser Systems defined by Lie Algebras}

As Olshanetsky and Perelomov [8,9]
realized very early on,
the Hamiltonian system (2.1) is only one
example of a whole series of Hamiltonian systems associated
with each simple Lie algebra.
More precisely, given any simple Lie algebra $\G$,
Olshanetsky and Perelomov [8] introduced the system
with Hamiltonian
$$
H(x,p)
={1\over 2}\sum_{i=1}^rp_i^2
-{1\over 2}
\sum_{\a\in{\cal R}(\G)}
m_{|\a|}^2\wp(\a\cdot x),
\eqno(2.5)
$$
where $r$ is the rank of $\G$, and ${\cal R}(\G)$
denotes the set of roots of $\G$. 
The $m_{|\a|}$ are mass parameters.
To preserve the invariance of the Hamiltonian (2.5)
under the Weyl group,
the parameters $m_{|\a|}$ depend only
on the length of $|\a|$ of the root
$\a$, and not on the root $\a$ itself.
In the case of $A_{N-1}=
SU(N)$,
it is common practice to use $N$ pairs of dynamical variables
$(x_i,p_i)$,
since the roots of $A_{N-1}$
lie conveniently on a hyperplane in ${\bf C}^N$.
The dynamics of the system are unaffected if we shift
all $x_i$ by a constant, and the number of degrees of freedom
is effectively $N-1=r$. Now the roots of $SU(N)$ are given
by $\a=e_i-e_j$, $1\leq i,j\leq N$, $i\not=j$. Thus
we recognize the original elliptic Calogero-Moser system
as the special case of (2.5) corresponding to $A_{N-1}$.
As in the original case, the elliptic systems (2.5)
admit rational and trigonometric limits.
Olshanetsky and Perelomov succeeded in constructing a
Lax pair for all these systems in the case of classical
Lie algebras, albeit without spectral parameter.

\bigskip
\noindent
{\bf Twisted Calogero-Moser Systems defined by Lie Algebras}

It turns out that the Hamiltonian systems (2.5) are not the
only natural extensions of the basic elliptic Calogero-Moser
system. 
A subtlety arises for simple Lie algebras $\G$ which are not
simply-laced, i.e., algebras which admit roots of uneven
length. This is the case for the algebras $B_n$, $C_n$, $G_2$,
and $F_4$ in Cartan's classification.
For these algebras, the following {\it twisted} elliptic
Calogero-Moser systems were introduced by the authors in [10,11]
$$
H_{\G}^{twisted}
=
{1\over 2}\sum_{i=1}^rp_i^2
-{1\over 2}
\sum_{\a\in{\cal R}(\G)}
m_{|\a|}^2
\wp_{\nu(\a)}(\a\cdot x).
\eqno(2.6)
$$
Here the function $\nu(\a)$ depends only on the length of the root $\a$.
If $\G$ is simply-laced, we set $\nu(\a)=1$ identically. Otherwise,
for $\G$ non simply-laced, we set
$\nu(\a)=1$ when $\a$ is a long root,
$\nu(\a)=2$ when $\a$ is a short root and $\G$ is one of the
algebras $B_n$, $C_n$, or $F_4$,
and $\nu(\a)=3$ when $\a$ is a short root
and $\G=G_2$.
The {\it twisted} Weierstrass function $\wp_{\nu}(z)$
is defined by
$$
\wp_{\nu}
(z)
=\sum_{\sigma=0}^{\nu-1}
\wp(z+2\omega_a{\sigma\over\nu}),
\eqno(2.7)
$$
where $\omega_a$ is any of the half-periods $\omega_1$,
$\omega_2$, or $\omega_1+\omega_2$.
Thus the twisted and untwisted Calogero-Moser systems coincide
for $\G$ simply laced. 
The original motivation for twisted
Calogero-Moser systems was based on
their scaling limits
(which will be discussed in the next section) [10,11].
Another motivation based on the symmetries of Dynkin diagrams
was proposed subsequently by Bordner, Sasaki, and Takasaki [20].

\bigskip
\centerline{\bf III. SCALING LIMITS OF CALOGERO-MOSER SYSTEMS}

\bigskip

\noindent
{\bf Results of Inozemtsev for $A_n$}

For the standard elliptic Calogero-Moser systems
corresponding to $A_{N-1}$, Inozemtsev [21,22] has shown in the
1980's that in the scaling limit
$$
\eqalignno{m&=Mq^{-{1\over 2N}},\ \ \ \ q\rightarrow0&(3.1)\cr
x_i&=X_i-2\omega_2{i\over N},
\
1\leq i\leq N &(3.2)\cr}
$$
where $M$ is kept fixed,
the elliptic $A_{N-1}$ Calogero-Moser Hamiltonian tends to
the following Hamiltonian 
$$
H_{Toda}
={1\over 2}\sum_{i=1}^Np_i^2
-{1\over 2}\big(\sum_{i=1}^{N-1}e^{X_{i+1}-X_i}+e^{X_1-X_N}\big)
\eqno(3.3)
$$
The roots $e_i-e_{i+1}$, $1\leq i\leq N-1$, and $e_N-e_1$
can be recognized as the simple roots of the
affine algebra $A_{N-1}^{(1)}$.
(For basic facts on affine algebras, we refer to [23]).
Thus (3.3) can be recognized as the Hamiltonian of the Toda system
defined by $A_{N-1}^{(1)}$.

\bigskip
\noindent
{\bf Scaling Limits based on the Coxeter Number}

The key feature of the above scaling limit is the collapse
of the sum over the
entire root lattice of $A_{N-1}$
in the Calogero-Moser Hamiltonian to the
sum over only simple roots
in the Toda Hamiltonian for the
Kac-Moody algebra $A_{N-1}^{(1)}$.
Our task is to extend this mechanism to general Lie algebras.
For this, we consider the following generalization
of the preceding scaling limit
$$
\eqalignno{
m&=Mq^{-{1\over 2}\d},&(3.4)\cr
x&=X-2\omega_2\d\rho^{\vee},&(3.5)\cr}
$$
Here $x=(x_i)$, $X=(X_i)$ and $\rho^{\vee}$
are $r$-dimensional vectors.
The vector $x$ is the dynamical
variable of the Calogero-Moser system.
The parameters $\d$ and $\rho^{\vee}$
depend on the algebra $\G$ and are yet to be chosen.
As for $M$ and $X$, they have the same interpretation as
earlier, namely as respectively the mass parameter
and the dynamical variables of the limiting system.
Setting $\omega_1=-i\pi$,
the contribution of each root $\a$ to the Calogero-Moser
potential can be expressed as
$$
m^2\wp(\a\cdot x)
=
{1\over 2}M^2
\sum_{n=-\infty}^{\infty}
{e^{2\d\omega_2}\over
{\rm ch}(\a\cdot x-2n\omega_2)-1}
\eqno(3.6)
$$
It suffices to consider positive roots $\a$.
We shall also assume that $0\leq \d\,\a\cdot\rho^{\vee}
\leq 1$. The contributions of the $n=0$ and $n=-1$
summands in (3.6) are proportional
to $e^{2\omega_2(\d-\d\,\a\cdot\rho^{\vee})}$
and $e^{2\omega_2(\d-1+\d\,\a\cdot\rho^{\vee})}$
respectively.
Thus the existence of a finite scaling limit requires
that
$$
\d\,\leq\d\,\a\cdot\rho^{\vee}\leq 1-\d.
\eqno(3.7)
$$
Let $\a_i$, $1\leq i\leq r$ be a basis of simple roots
for $\G$. If we want all simple roots $\a_i$
to survive in the limit, we must require that
$$
\a_i\cdot\rho^{\vee}=1,\ \
1\leq i\leq r.
\eqno(3.8)
$$
This condition characterizes the vector $\rho^{\vee}$
as the {\it level vector}.
Next, the second condition in (3.7)
can be rewritten as $\d\{1+max_{\a}\,(\a\cdot\rho^{\vee})\}
\leq 1$. But
$$
h_{\G}=1+max_{\a}\,(\a\cdot\rho^{\vee})
\eqno(3.9)
$$
is precisely the Coxeter number of $\G$,
and we must have $\d\leq {1\over h_{\G}}$.
Thus when $\d<{1\over h_{\G}}$,
the contributions of all the roots except
for the simple roots of $\G$ tend to $0$.
On the other hand, when $\d={1\over h_{\G}}$,
the highest root $\a_0$ realizing the maximum over
$\a$ in (3.9) survives.
Since $-\a_0$ is the additional
simple root for the affine Lie algebra
$\G^{(1)}$, we arrive in this way at the following theorem,
which was proved in [11]

\bigskip
\noindent
{\bf Theorem 1}.
{\it Under the limit (3.4-3.5), with $\d={1\over h_{\G}}$,
and $\rho^{\vee}$ given by the
level vector,
the Hamiltonian of the elliptic Calogero-Moser system
for the simple Lie algebra $\G$
tends to the Hamiltonian of the Toda system
for the affine Lie algebra $\G^{(1)}$.}

\bigskip

\noindent
{\bf Scaling Limit based on the Dual Coxeter Number}

If the Seiberg-Witten spectral curve of the $\N=2$
supersymmetric gauge theory with a hypermultiplet in
the adjoint representation is to be realized as
the spectral curve for a Calogero-Moser system,
the parameter $m$ in the Calogero-Moser system
should correspond to the mass of the hypermultiplet.
In the gauge theory, the dependence of the coupling
constant on the mass $m$ is given by
$$
\tau={i\over 2\pi}h_{\G}^{\vee}{\rm ln}\,{m^2\over M^2}
\qquad
\Longleftrightarrow
\qquad
m=Mq^{-{1\over 2h_{\G}^{\vee}}}
\eqno(3.10)
$$
where $h_{\G}^{\vee}$ is the quadratic Casimir of the
Lie algebra $\G$. This shows that the correct physical
limit, expressing the decoupling of the hypermultiplet
as it becomes infinitely massive,
is given by (3.4), but with $\d={1\over h_{\G}^{\vee}}$.
To establish a closer parallel with our preceding discussion,
we recall that the quadratic Casimir $h_{\G}^{\vee}$
coincides with the {\it dual Coxeter number} of $\G$,
defined by
$$
h_{\G}^{\vee}=1+max_{\a}\,(\a^{\vee}\cdot\rho),
\eqno(3.11)
$$
where $\a^{\vee}={2\a\over\a^2}$ is the coroot associated
to $\a$, and $\rho={1\over 2}\sum_{\a>0}\a$
is the well-known Weyl vector.

For simply laced Lie algebras $\G$ (ADE algebras),
we have $h_{\G}=h_{\G}^{\vee}$, and the preceding scaling limits
apply. However, for non simply-laced algebras
($B_n$, $C_n$, $G_2$, $F_4$), 
we have $h_{\G}>h_{\G}^{\vee}$,
and our earlier considerations show that the untwisted
elliptic Calogero-Moser Hamiltonians do not tend to
a finite limit under (3.10), $q\to 0$,
$M$ is kept fixed.
This is why the twisted Hamiltonian systems (2.6)
have to be introduced. The twisting produces precisely
to an improvement in the asymptotic behavior
of the potential which allows a finite, non-trivial limit.
More precisely,
we can write
$$
m^2\wp_{\nu}(x)
=
{c_{\nu}\over 2}
\sum_{n=-\infty}^{\infty}
{m^2\over {\rm ch}\,\nu(x-2n\omega_2)-1},
\eqno(3.12)
$$
where $c_{\nu}=\nu^2$.
Setting $x=X-2\omega_2\d^{\vee}\rho$, we
obtain the following asymptotics
$$
m^2\wp_{\nu}(x)
=c_{\nu}M^2
\cases{e^{-2\omega_2(\d^{\vee}\a^{\vee}\cdot\rho-\d^{\vee})-\a^{\vee}\cdot X}
+e^{-2\omega_2(1-\d^{\vee}\a^{\vee}\cdot\rho-\d^{\vee})+\a^{\vee}\cdot X},
&if $\a$ is long;\cr
e^{-2\omega_2(\d^{\vee}\a^{\vee}\cdot\rho-\d^{\vee})-\a^{\vee}\cdot X},
&if $\a$ is short.\cr}
\eqno(3.13)
$$
This leads to the following theorem [11]

\bigskip
\noindent
{\bf Theorem 2}.
{\it 
Under the limit $x=X+2\omega_2{1\over h_{\G}^{\vee}}\rho$,
$m=Mq^{-{1\over 2h_{\G}^{\vee}}}$,
with $\rho$ the Weyl vector and $q\to 0$,
the Hamiltonian of the twisted elliptic Calogero-Moser system
for the simple Lie algebra $\G$
tends to the Hamiltonian of the Toda system
for the affine Lie algebra $(\G^{(1)})^{\vee}$.}

\bigskip

This suggests that the twisted Calogero-Moser system
is the integrable model solving the N=2 supersymmetric
gauge theory with gauge algebra $\G$ since,
in view of the work of Martinec and Warner [24],
it is the Toda system for $(\G^{(1)})^{\vee}$
which solves the corresponding pure Yang-Mills theory.

\bigskip

So far we have discussed only the scaling limits of the Hamiltonians.
However, similar arguments show that the Lax pairs constructed below
also have finite, non-trivial scaling limits whenever
this is the case for the Hamiltonians.
The spectral parameter $z$ should scale as $e^z=Zq^{1\over 2}$,
with $Z$ fixed. The parameter $Z$ can be identified
with the loop group parameter
for the resulting affine Toda system.

\bigskip

\centerline{\bf IV. LAX PAIRS FOR CALOGERO-MOSER SYSTEMS}

\bigskip

\noindent
{\bf The General Ansatz}

Let the rank of $\G$ be $n$,
and $d$ be its dimension.
Let $\L$ be a representation of $\G$ of dimension $N$,
of weights $\l_I$, $1\leq I\leq N$. Let $u_I\in {\bf C}^N$
be the weights of the fundamental
representation of $GL(N,{\bf C})$. Project
orthogonally the $u_I$'s onto the $\l_I$'s as
$$
su_I=\l_I+u_I,
\ \
\l_I\perp v_J.
\eqno(4.1)
$$
It is easily verified that $s^2$ is the second Dynkin index.
Then
$$
\a_{IJ}=\l_I-\l_J
\eqno(4.2)
$$
is a weight of $\L\otimes\L^*$ associated to the
root $u_I-u_J$ of $GL(N,{\bf C})$. The Lax pairs
for both untwisted and twisted Calogero-Moser systems
will be of the form
$$
L=P+X,
\ \
M=D+X,
\eqno(4.3)
$$
where the matrices $P,X,D$, and $Y$ are given by
$$
X=\sum_{I\not=J}C_{IJ}\P_{IJ}(\a_{IJ},z)E_{IJ},
\ \ \
Y=\sum_{I\not=J}C_{IJ}\P'_{IJ}(\a_{IJ},z)E_{IJ}
\eqno(4.4)
$$
and by
$$
P=p\cdot h,
\ \ \ \
D=d\cdot (h\oplus\tilde h)+\Delta.
\eqno(4.5)
$$
Here $h$ is in a Cartan subalgebra ${\cal H}_{\G}$
for $\G$, $\tilde h$
is in the Cartan-Killing orthogonal
complement of ${\cal H}_{\G}$
inside a Cartan subalgebra ${\cal H}$ for $GL(N,{\bf C})$,
and $\Delta$ is in the centralizer of ${\cal H}_{\G}$
in $GL(N,{\bf C})$.
The functions $\P_{IJ}(x,z)$ and the coefficients
$C_{IJ}$ are yet to be determined.
We begin by stating the necessary and sufficient
conditions for the pair $L(z)$, $M(z)$ of (4.1)
to be a Lax pair for the
(twisted or untwisted) Calogero-Moser
systems. For this, it is convenient to
introduce the following notation
$$
\eqalignno{
\P_{IJ}&=\P_{IJ}(\a_{IJ}\cdot x)\cr
\wp_{IJ}'
&=\P_{IJ}(\a_{IJ}\cdot x,z)\P_{JI}'(-\a_{IJ}\cdot x,z)
-\P_{IJ}(-\a_{IJ}\cdot x,z)
\P_{JI}'(\a_{IJ}\cdot x,z).
&(4.6)
\cr}
$$
Then the Lax equation $\dot L(z)
=[L(z),M(z)]$ implies the
Calogero-Moser system if and only
if the following three identities are satisfied
$$
\sum_{I\not=J}C_{IJ}C_{JI}\wp_{IJ}'\a_{IJ}
=
s^2\sum_{\a\in {\cal R}(\G)}
m_{|\a|}^2\wp_{\nu(\a)}(\a\cdot x)
\eqno(4.7)
$$
$$
\sum_{I\not=J}C_{IJ}C_{JI}
\wp_{IJ}'(v_I-v_J)
=0
\eqno(4.8)
$$
$$
\eqalignno{
\sum_{K\not= I,J}
C_{IK}C_{KJ}(\P_{IK}\P_{KJ}'-\P_{IK}'\P_{KJ})
&=
sC_{IJ}\P_{IJ}d\cdot (v_I-v_J)
+
\sum_{K\not= I,J}
\Delta_{IJ}C_{KJ}\P_{KJ}\cr
&\qquad\qquad\qquad
-
\sum_{K\not= I,J}
C_{IK}\P_{IK}\Delta_{KJ}
&(4.9)\cr}
$$
\noindent
The following theorem was established in [10]:

\bigskip

\noindent
{\bf Theorem 3}. {\it A representation $\Lambda$, functions
$\Phi_{IJ}$, and coefficients $C_{IJ}$ with a spectral parameter $z$
satisfying (4.7-4.9) can be found for all twisted and untwisted elliptic
Calogero-Moser systems associated with a simple Lie algebra
$\G$, except possibly in the case of twisted $G_2$.
In the case of $E_8$, we have to assume the existence of a
$\pm1$ cocycle.}   

\bigskip

\noindent
{\bf Lax Pairs for Untwisted Calogero-Moser Systems}

We now describe some important
features of the Lax pairs we obtain in this manner.

\bigskip

$\bullet$ In the case of the {\it untwisted} Calogero-Moser systems,
we can choose $\P_{IJ}(x,z)=\P(x,z)$,
$\wp_{IJ}(x)=\wp(x)$ for all $\G$.

\medskip

$\bullet$ $\Delta=0$ for all $\G$, except for $E_8$.

\medskip

$\bullet$ For $A_n$, the Lax pair (2.2-2.3) corresponds
to the choice of the fundamental representation for $\L$.
A different Lax pair can be found by taking
$\L$ to be the antisymmetric representation.
 
\medskip

$\bullet$ For the $BC_n$ system, the Lax pair is
obtained by imbedding $B_n$ in $GL(N,{\bf C})$
with $N=2n+1$. When $z=\omega_a$ (half-period),
the Lax pair obtained this way reduces to
the Lax pair obtained by Olshanetsky and Perelomov [8,9].

\medskip 

$\bullet$ For the $B_n$ and $D_n$ systems,
additional Lax pairs with spectral parameter
can be found by taking $\L$ to be
the spinor representation.

\medskip

$\bullet$ For $G_2$,
a first Lax pair with spectral parameter can be
obtained by the above construction
with $\L$ chosen to be the ${\bf 7}$ of $G_2$.
A second Lax pair with spectral parameter
can be obtained by restricting the {\bf 8}
of $B_3$ to the ${\bf 7}\oplus{\bf 1}$
of $G_2$.

\medskip

$\bullet$ For $F_4$, a Lax pair can be obtained by
taking $\L$ to be the ${\bf 26}\oplus{\bf 1}$
of $F_4$, viewed as the restriction of
the {\bf 27} of $E_6$ to its $F_4$ subalgebra.

\medskip

$\bullet$ For $E_6$, $\L$ is the {\bf 27} representation.

\medskip

$\bullet$ For $E_7$, $\L$ is the {\bf 56} representation.

\medskip

$\bullet$ For $E_8$, a Lax pair with spectral parameter
can be constructed with $\L$ given by the {\bf 248} representation,
if coefficients $c_{IJ}=\pm 1$ exist
with the following cocycle conditions
$$
\eqalignno{
c(\lambda,\lambda-\d)c(\lambda-\d,\mu)=&
c(\lambda,\mu+\d)c(\mu+\d,\mu)\cr
&{\rm \ when\ \d\cdot\lambda=-\d\cdot\mu=1,
\
\lambda\cdot\mu=0}\cr
c(\l,\mu)c(\l-\d,\mu)=&c(\l,\l-\d)\cr
& {\rm \ when\ \d\cdot\l=\l\cdot\mu=1,
\
\d\cdot\mu=0}\cr
c(\l,\mu)c(\l,\l-\mu)=&
-c(\l-\mu,-\mu)\cr
&{\rm \ when\ \l\cdot\mu=1}.
&(4.10)
\cr}
$$
The matrix $\Delta$ in the Lax pair is then the $8\times 8$ matrix
given by
$$
\eqalignno{
\Delta_{ab}=&
\sum_{\d\cdot\b_a=1\atop \d\cdot\b_b=1}
{m_2\over 2}
\big(c(\b_a,\d)c(\d,\b_b)
+
c(\b_a,\b_a-\d)c(\b_a-\d,\b_b)\big)
\wp(\d\cdot x)\cr
&
-\sum_{\d\cdot\b_a=1\atop \d\cdot\b_b=-1}
{m_2\over 2}
\big(c(\b_a,\d)c(\d,\b_b)
+
c(\b_a,\b_a-\d)c(\b_a-\d,\b_b)\big)
\wp(\d\cdot x)\cr
\Delta_{aa}=&
\sum_{\b_a\cdot\d=1}
m_2\wp(\d\cdot x)
+2m_2\wp(\b_a\cdot x),
&(4.11)
\cr}
$$
where $\b_a$, $1\leq a\leq 8$, is a maximal set of 8 mutually
orthogonal roots.

\medskip

We note that recently Lax pairs of root type have been
considered [20,25] which correspond, in the above Ansatz (4.3-5),
to $\Lambda$ equal to the adjoint representation of $\G$
and the coefficients $C_{IJ}$ vanishing for $I$ or $J$ 
associated with zero weights.
This construction yields another Lax pair for the case $E_8$.
Spectral curves for certain gauge theories with matter
in the adjoint representation have also
been proposed in [26] and [27], based on branes and M-theory.
\bigskip

\noindent
{\bf Lax Pairs for Twisted Calogero-Moser Systems}

Recall that the twisted and untwisted Calogero-Moser systems
differ only for non-simply laced Lie algebras, namely
$B_n$, $C_n$, $G_2$ and $F_4$.
These are the only algebras we discuss in
this paragraph.
The construction (4.3-4.9) gives then
Lax pairs for all of them, 
with the possible exception of twisted $G_2$.
Unlike the case of untwisted Lie algebras however,
the functions $\P_{IJ}$ have to be chosen
with care, and differ for each algebra.
More specifically,

\medskip

$\bullet$ For $B_n$, the Lax pair is of dimension $N=2n$,
admits two independent couplings $m_1$ and $m_2$,
and
$$
\P_{IJ}(x,z)
=
\cases{
\P(x,z), &if $I-J\not= 0,\pm n$\cr
\P_2({1\over 2}x,z), &if $I-J=\pm n$\cr}.
\eqno(4.12)
$$
Here a new function $\P_2(x,z)$ is defined by
$$
\P_2({1\over 2}x,z)
={\P({1\over 2}x,z)\P({1\over 2}x+\omega_1,z)
\over
\P(\omega_1,z)}
\eqno(4.13)
$$

$\bullet$ For $C_n$, the Lax pair is of dimension $N=2n+2$,
admits one independent coupling $m_2$,
and
$$
\P_{IJ}(x,z)
=
\P_2(x+\omega_{IJ},z),
$$
where $\omega_{IJ}$ are given by
$$
\omega_{IJ}
=
\cases{0, &if $I\not=J=1,2,\cdots,2n+1$;\cr
\omega_2, &if $1\leq I\leq 2n,\ J=2n+2$;\cr
-\omega_2, &if $1\leq J\leq 2n,\ I=2n+2$.\cr}
\eqno(4.14)
$$

$\bullet$ For $F_4$, the Lax pair is of dimension $N=24$,
two independent couplings $m_1$ and $m_2$,
$$
\P_{\l\mu}(x,z)
=
\cases{\P(x,z), &if $\l\cdot\mu=0$;\cr
\P_1(x,z), &if $\l\cdot\mu={1\over 2}$;\cr
\P_2({1\over 2}x,z), &if $\l\cdot\mu=-1$.\cr}
\eqno(4.15)
$$
where the function $\P_1(x,z)$ is defined by
$$
\P_1(x,z)
=
\P(x,z)
-
e^{\pi i\zeta(z)+\eta_1z}
\P(x+\omega_1,z)
\eqno(4.16)
$$
Here it is more convenient to label
the entries of the Lax pair directly by the weights
$\lambda=\lambda_I$ and
$\mu=\lambda_J$ instead of $I$ and $J$.

$\bullet$ For $G_2$, candidate Lax pairs
can be defined in the {\bf 6} and {\bf 8}
representations of $G_2$, but it is still unknown whether
elliptic functions $\P_{IJ}(x,z)$
exist which satisfy the required identities.

\bigskip
\centerline{\bf V. CALOGERO-MOSER AND SPIN CALOGERO-MOSER}
\centerline{\bf SPECTRAL CURVES}

\bigskip

A Lax pair $L(z),M(z)$ with spectral parameter gives rise to
a spectral curve $\Gamma$ defined by
$$
\Gamma=\{(k,z);\ R(k,z)\equiv det(kI-L(z))=0\}
\eqno(5.1)
$$
Since the matrix $L(z)$ is expressed in terms of the dynamical
variables of the Calogero-Moser system, the family of spectral
curves $\Gamma$ can be parametrized by constants of motion
of the system. However, to make contact with supersymmetric gauge
theories, it is important to find parametrizations
of the spectral curves in terms of the order parameters
of the gauge theory. This problem was solved for the
$A_{N-1}$ Calogero-Moser systems in [3]. Here we extend the solution
given there to the more general class of $SL(N,{\bf C})$
{\it spin Calogero-Moser systems}.

\bigskip

The $SL(N,{\bf C})$ spin Calogero-Moser system introduced
in [15] is the system with Hamiltonian
$$
H={1\over 2}\sum_{i=1}^Np_i^2
-{1\over 2}m^2\sum_{i\not=j}(b_i^{\dagger}a_j)(b_j^{\dagger}a_i)
V(x_i-x_j)
\eqno(5.2)
$$
The terms $a_i=(a_i)_{\alpha}$, $b_i=(b_i)^{\alpha}$ are respectively 
$l$-dimensional vectors and $l$-dimensional covectors,
and $b_i^{\dagger}a_j$ is their scalar product.
The system (5.2) admits a Lax pair $L(z)$, $M(z)$ which is a generalization
of (2.2). In particular, $L(z)$ is given by
$$
L_{ij}(z)=p_i\d_{ij}-m(1-\d_{ij})f_{ij}\P(x_i-x_j,z)
\eqno(5.3)
$$
with
$$
f_{ij}=b_i^{\dagger}a_j,\ \ \ f_{ii}=m.
\eqno(5.4)
$$
Krichever et al. have shown that the corresponding family of spectral
curves $\Gamma$ is a $Nl-{1\over 2}l(l-1)$-dimensional 
family of Riemann surfaces of genus
$g=Nl+1-{1\over 2}l(l+1)$. 
The defining equation $R(k,z)=0$ can be expressed in the form
$$
R(k,z)
=k^N+\sum_{i=0}^{N-1}r_i(z)k^i
$$
where $r_i(z)$, $0\leq i\leq N-1$, is an elliptic function 
with a pole of order $N-i$.
Since elliptic functions can be
expanded linearly in terms of $\wp(z)$ and $\wp'(z)$,
the family of spectral curves $\Gamma$
can be parametrized by
the coefficients of $r_i(z)$ in such an expansion.
The number of these coefficients exceeds $Nl-{1\over 2}l(l-1)$
however,
and Krichever et al. [15] show that the correct number
of parameters can be obtained by imposing
linear constraints on the coefficients.

\bigskip

We present now a different parametrization of the spectral curves
of the spin Calogero-Moser systems, motivated by the order parameters
of $N=2$ supersymmetric $SU(N)$ gauge theories.
As in [3], we introduce the functions $h_n(z)$ by
$$
h_n(z)={\partial_z^n\theta_1({z\over 2\omega_1}|\tau)
\over\theta_1({z\over 2\omega_1}|\tau)},
\ \
n\in{\bf N}.
\eqno(5.5)
$$
and set
$$
f(k,z)=R(k+mh_1(z),z).
\eqno(5.6)
$$

\bigskip
\noindent
{\bf Theorem 4}. {\it The function $f(k,z)$ can be expressed as
$$
f(k,z)
=
\sum_{p=1}^l
\partial_z^{p-1}\big(
{\theta_1({1\over 2\omega_1}(z-m{\partial\over\partial k})|\tau)
\over
\theta_1({1\over 2\omega_1}|\tau)}
H_p(k)\big)
\eqno(5.7)
$$
where $H_p(k)$ is a polynomial 
in $k$ of degree $N-p+1$, for $1\leq p\leq l$.} 

\bigskip
The polynomial $H_1(k)$
is monic because $R(k,z)$ and $f(k,z)$
are. As for the polynomials $H_p(k)$
with $p>1$,
their terms of order $k^0$
do not contribute in (5.7)
and may be taken to be $0$.
Thus we note that the total number of parameters for the $l$ monic polynomials
$H_p(k)$ is $\sum_{p=1}^l(N-p+1)=lN-{1\over 2}l(l-1)$,
which is indeed the dimension of the family
of spectral curves $\Gamma$ for the $SL(N,{\bf C})$ spin
Calogero-Moser system.

\bigskip
\noindent
{\it Proof of Theorem 4}. It is easily seen
from the transformation
properties of $h_1(z)$ that the transformation properties
for $f(k,z)$ are
$$
\eqalignno{
f(k,z+2\omega_1)=&f(k,z)\cr
f(k,z+2\omega_2)=&f(k-\beta m,z),\ \ \beta=-{i\pi\over\omega_1} 
&(5.8)\cr}
$$
Furthermore, the function $f(k,z)$ has poles at $z=0$, with the
residue a polynomial in $k$ of degree $N-p$ at a pole of order $p$.
Now the functions $h_n(z)$ satisfy the monodromy conditions
$$
\eqalignno{
h_n(z+2\omega_1)=&h(z)\cr
h_n(z+2\omega_2)=&\sum_{p=0}^n\pmatrix{n\cr p\cr}\beta^{n-p}h_p(z).
&(5.9)
\cr}
$$
It follows that the monodromies for their derivatives are
$$
\eqalignno{
\partial_z^sh_n(z+2\omega_1)=&\partial_z^s h(z)\cr
\partial_z^sh_n(z+2\omega_2)=&\sum_{p=1}^n\pmatrix{n\cr p\cr}\beta^{n-p}
\partial_z^sh_p(z).
&(5.10)
\cr}
$$
(The $p=0$ term in the second identity does not contribute
since $h_0(z)=1$.) Note also that $\partial_zh_1(z)
=\partial_z^2{\rm log}\,\theta_1({z\over 2\omega_1}|\tau)
=-4\omega_1^2\wp(z)$ is doubly periodic.
Thus we may set
$$
f(k,z)
=
\sum_{p=1}^l
\sum_{n=0}^{N-p+1}
Q_{p,N-p+1-n}(k)
\partial_z^{p-1}h_n(z).
\eqno(5.11)
$$
Next, we translate the monodromy transformations for $f(k,z)$
in terms of the polynomials $Q_{p,N-p+1-n}(k)$. We may write
$$
\eqalignno{
f(k,z)=&\sum_{n=0}^Nh_n(z)Q_{1,N-n}(k)\cr
&+\sum_{p=2}^l\sum_{n=1}^{N-p+1}\partial_z^{p-1}h_n(z)Q_{p,N-p+1-n}(k)\cr
f(k,z+2\omega_2)
=&\sum_{n=0}^N\sum_{s=0}^n\pmatrix{n\cr s\cr}
\b^{n-s}h_1(z)Q_{1,N-n}(k)\cr
&+\sum_{p=2}^l\sum_{n=1}^{N-p+1}
\sum_{s=1}^n\pmatrix{n\cr s\cr}
\b^{n-s}\partial_z^{p-1}h_s(z)
Q_{p,N-p+1-n}(k).
&(5.12)
\cr}
$$
But the functions $h_n(z)$, $0\leq n\leq N$, and
$\partial_z^{p-1}h_n(z)$, $1\leq n\leq N$, $2\leq p\leq l$,
are linearly independent. Thus we may equate coefficients
and obtain for $Q_{1,N-s}(k)$ the relation
$$
Q_{1,N-s}
=
\sum_{n=0}^N\pmatrix{n\cr s\cr}
\b^{n-s}Q_{1,N-n}(k)\eqno(5.13)
$$
Changing $N-s\rightarrow p$ and $N-n\rightarrow n$,
this can be rewritten as
$$
Q_{1,p}(k-\b m)
=
\sum_{n=0}^N\pmatrix{N-n\cr p-n\cr}
\b^{p-n}Q_{1,n}(k)
\eqno(5.14)
$$
This is a relation of the form studied in [5],(3.9).
We recall briefly the argument: the equation (5.14)
is equivalent to the equation
$H(t+\beta,k+\beta m)=H(t,k)$ where $H(t,k)=\sum_{p=0}^Nt^{N-p}Q_{1,p}(k)$
is the generating function. Since $H(t,k)$ is a polynomial
in both $t$ and $k$, this means that $H(t,k)=H(0,k-tm)$
depends only on $k-tm$. Setting $H_1(k)=H(0,k)$,
it follows easily that
$$
Q_{1,N-n}(k)={(-m)^n\over n!}H_1^{(n)}(k).
\eqno(5.15)
$$
Next, we solve for the higher order terms $Q_{p,N-p+1-s}(k)$.
They satisfy
$$
Q_{p,N-p+1-s}
(k-\b m)
=
\sum_{n=1}^{N-p+1}
\pmatrix{n\cr s\cr}
\b^{n-s}
Q_{p,N-p+1-n}(k)
\eqno(5.16)
$$
This is again a relation of the form (5.14), with $N$ replaced by
$N-p+1$. Thus there is again a polynomial $H_p(k)$,
of degree $N-p+1$ so that
$$
Q_{p,N-p+1}(k)
=
{(-m)^s\over s!}H_p^{(s)}(k).
\eqno(5.17)
$$
Substituting in (5.11), and noting that
$$
{\theta_1({1\over 2\omega_1}(z-m{\partial\over\partial k})|\tau)
\over
\theta_1({z\over 2\omega_1}|\tau)}
=
\sum_{n=0}^{\infty}
h_n(z){(-m)^n\over n!}({\partial\over\partial k})^n,
\eqno(5.18)
$$
we obtain the desired expression (5.7).

\bigskip

Evidently, the coefficients of the polynomials $H_p(k)$
(or equivalently, their zeroes) are integrals of motion
of the spin Calogero-Moser system.
It would be valuable to express them directly in terms
of the dynamical variables $(p_i,x_i)$ of the system.
For the $SU(N)$ Calogero-Moser system,
this problem was solved in [28].

\bigskip
Finally, we would like to 
note also that in the simpler case of the $SU(N)$ Calogero-Moser
system, an alternative derivation of the parametrization in [3] is now
available [29]. It would be interesting to explore 
also generalizations of this new derivation.

\vfill\break

\centerline{\bf REFERENCES}
\bigskip

\item{[1]} Donagi, R. and E. Witten,
``Supersymmetric Yang-Mills and integrable systems",
Nucl. Phys. {\bf B 460} (1996) 288-334, hep-th/9510101.

\item{[2]} Gorsky, A. and N. Nekrasov, ``Elliptic Calogero-Moser
systems from two-dimensional current algebra", hep-th/9401021;
N. Nekrasov, ``Holomorphic bundles and many-body systems",
Comm. Math. Phys. {\bf 180} (1996) 587; Martinec, E.,
``Integrable structures in supersymmetric gauge
and string theory", hep-th/9510204.

\item{[3]} D'Hoker, E. and D.H. Phong,
``Calogero-Moser systems in $SU(N)$ Seiberg-Witten theory",
Nucl. Phys. {\bf B 513} (1998) 405-444, hep-th/9709053.

\item{[4]} Krichever, I.M. and D.H. Phong,
``Symplectic forms in the theory of solitons",
hep-th/9708170, to appear in Surveys in Differential Geometry,
Vol. III.

\item{[5]} Lerche, W.,
``Introduction to Seiberg-Witten theory
and its stringy origins",
Proceedings of the {\it Spring School and Workshop
on String Theory}, ICTP, Trieste (1996),
hep-th/9611190, Nucl. Phys. Proc.
Suppl. {\bf B 55} (1997) 83.

\item{[6]} Marshakov, A.,
``On integrable systems and supersymmetric gauge
theories", Theor. Math. Phys. {\bf 112} (1997)
791-826, hep-th/9702083.

\item{[7]} Marshakov, A., A. Mironov, and A. Morozov,
``WDVV-like equations in N=2 SUSY Yang-Mills theory",
Phys. Lett. {\bf B 389} (1996) 43, hep-th/9607109;
``More evidence for the WDVV equations
in N=2 SUSY Yang-Mills theory",
hep-th/9701123.

\item{[8]} Olshanetsky, M.A. and A.M. Perelomov,
``Completely integrable Hamiltonian systems
connected with semisimple Lie algebras",
Inventiones Math. {\bf 37} (1976) 93-108.

\item{[9]} Olshanetsky, M.A. and A.M. Perelomov,
``Classical integrable finite-dimensional
systems related to Lie algebras",
Phys. Rep. {\bf 71 C} (1981) 313-400.

\item{[10]} D'Hoker, E. and D.H. Phong,
``Calogero-Moser Lax pairs with spectral parameter
for general Lie algebras", Nucl. Phys. {\bf B 530} (1998)
537-610, hep-th/9804124.

\item{[11]} D'Hoker, E. and D.H. Phong,
``Calogero-Moser and Toda systems for twisted and
untwisted affine Lie algebras",
Nucl. Phys. {\bf B 530} (1998) 611-640, hep-th/9804125.

\item{[12]} D'Hoker, E. and D.H. Phong,
``Spectral curves for super Yang-Mills with adjoint
hypermultiplet for general Lie algebras",
Nucl. Phys. {\bf B 534} (1998) 697-719, hep-th/9804126.

\item{[13]} Krichever, I.M. and A. Zabrodin,
``Spin generalizations of the spin Ruijsenaars-Schnei\-der
model, non-abelian 2D Toda chain,
and representations of the Sklyanin algebra",
hep-th/9505039.

\item{[14]} Ruijsenaars, S.N.M. and H. Schneider,
``A new class of integrable systems and its relation
to soliton equations",
Ann. Phys. (NY) {\bf 170} (1986) 370-405;
S.N.M Ruijsenaars, Comm. Math. Phys. {\bf 110} (1987) 191; see
also
A. Gorsky and N. Nekrasov, ``Relativistic Calogero-Moser
Model as a gauged WZW model", Nucl. Phys. {\bf B436}
(1995) 582, hep-th/9401017; H.W. Braden and R. Sasaki,
Prog. Theor. Phys. {\bf 97} (1997) 1003.

\item{[15]} Krichever,I.M., O. Babelon, E. Billey, and M. Talon,
``Spin generalization of the Caloge\-ro-Moser system and the
matrix KP equation",
Amer. Math. Soc. Transl. {\bf 170} (1995) 83-119;
J.A. Minahan and A.P. Polychronakos, ``Integrable systems for
particles with internal degrees of freedom", Phys. Lett. {\bf B302}
(1993) 265; ``Interacting Fermion Systems from Two Dimensional QCD", 
Phys. Lett. {\bf B326} (1994) 288, hep-th/9309044;
A.P. Polychronakos, ``Exchange Operator Formalism for Integrable
Systems of Particles", Phys. Rev. Lett. {\bf 69} (1992) 703,
hep-th/9202057;
``Generalized Statistics in one dimension", hep-th/9902157.

\item{[16]} Calogero, F.,
``Exactly solvable one-dimensional
many-body problems",
Lett. Nuovo Cim. {\bf 13} (1975)
411-416.

\item{[17]} Moser, J.,
``Three integrable Hamiltonian systems connected with
isospectral deformations",
Advances Math. {\bf 16} (1975) 197.

\item{[18]} Krichever, I.M.,
``Elliptic solutions of the Kadomtsev-Petviashvili equation
and integrable systems of particles",
Funct. Anal. Appl. {\bf 14} (1980) 282-290.

\item{[19]} Braden, H.W. and V.M. Buchstaber,
``The general analytic solution
of a functional equation of addition type",
Siam J. Math. anal. {\bf 28} (1997) 903-923.

\item{[20]} Bordner, A., R. Sasaki, and K. Takasaki,
``Calogero-Moser systems II: symmetries and
foldings", hep-th/9809068; A. Bordner and R. Sasaki, 
``Calogero-Moser systems III: Elliptic Potentials 
and Twisting, hep-th/9812232.

\item{[21]} Inozemtsev, I.,
``Lax representation
with spectral parameter on a torus for integrable
particle systems",
Lett. Math. Phys. {\bf 17} (1989) 11-17.

\item{[22]} Inozemtsev, I.,
``The finite Toda lattices",
Comm. Math. Phys. {\bf 121} (1989) 628-638.

\item{[23]} Goddard, P. and D. Olive,
``Kac-Moody and Virasoro algebras in relation to
quantum physics", International J. Mod. Phys. A, Vol. I (1986)
303-414.

\item{[24]} Martinec, E. and N. Warner,
``Integrable systems and supersymmetric gauge theories",
Nucl. Phys. {\bf B 459} (1996) 97-112, hep-th/9509161.

\item{[25]} Bordner, A., E. Corrigan, and R. Sasaki,
``Calogero-Moser systems: a new formulation",
hep-th/9805106.

\item{[26]} Uranga, A.M.,
``Towards mass deformed N=4 $SO(N)$ and $Sp(K)$ gauge
theories from brane configurations",
Nucl. Phys. {\bf B 526} (1998) 241-277,
hep-th/9803054.

\item{[27]} Yokono, T.,
``Orientifold four plane
in brane configurations
and N=4 $USp(2N)$
and $SO(2N)$ theory",
Nucl. Phys. {\bf B 532} (1998) 210-226,
hep-th/9803123.

\item{[28]} D'Hoker, E. and D.H. Phong,
``Order parameters, free fermions, and
conservation laws for Calogero-Moser systems",
hep-th/9808156, to appear in Asian J. Math.

\item{[29]} Vaninsky, K.,
``On explicit parametrization of spectral curves
for Moser-Calogero particles and its applications",
December 1998 preprint.
\item{[30]} Nekrasov, N., Nucl. Phys. {\bf B531} (1998) 323;
H.W. Braden, A. Marshakov, A. Mironov and A. Morozov,
``The Ruijsenaars-Schneider Model in the Context of
Seiberg-Witten Theory", hep-th/9902205.

\end